\documentstyle[twocolumn,aps]{revtex}
\begin{document}
\draft
\title{\bf {Heat Capacity of Mesoscopic Superconducting Disks}}
\author{P. Singha Deo$^1$\cite{eml}, J. P. Pekola$^2$, and M. Manninen$^2$} 
\address{$^1$S.N. Bose National Centre for Basic Sciences, J.D. Block,
Sector 3, Salt Lake City, Calcutta 91, India.\\
$^2$Department of Physics, University of Jyv{\"a}skyl{\"a},
P.O.Box 35, 40351 Jyv{\"a}skyl{\"a}, Finland.}
\maketitle
\begin{abstract}
We study the heat capacity of isolated giant vortex states, which
are good angular momentum ($L$) states, in a mesoscopic
superconducting disk using the Ginzburg-Landau (GL) theory. At small 
magnetic fields
the $L$=0 state qualitatively behaves like the bulk sample
characterized by a discontinuity in heat capacity at $T_c$. As the
field is increased the discontinuity slowly
turns into a continuous change which is a
finite size effect. The
higher $L$ states show a continuous change in heat capacity at $T_c$ at
all fields. We also show that for these higher $L$ states, the
behavior of the peak position with change in field is related to
the paramagnetic Meissner effect (irreversible)
and can lead to an unambiguous observation
of positive magnetization in mesoscopic superconductors.
\end{abstract}
\narrowtext

Superconducting state in mesoscopic samples can have  strikingly
different properties as compared to the corresponding
bulk samples because by definition [1-10], the intrinsic
length scales in them, 
like the penetration depth and the coherence length
are of the order of the sample dimensions.
Essentially, in a magnetic field a bulk type II superconductor can
exist either in the Meissner state (macroscopic condensation of the
Cooper pairs to the zero momentum state) or it can exist in a mixed
state of higher momenta with infinite components (when there are
vortices in the sample). In the presence of a boundary there can be
another state called the giant vortex state wherein a very thin
region near the surface is superconducting \cite{boo}.  This edge 
state in
bulk samples affects resistivity  measurements because contacts
applied at the edges of the  superconductor are shorted by it.
Therefore, resistivity measurements are useful to determine the
magnetic field to nucleate  the giant vortex states. Otherwise these
edge states are not capable of affecting the thermodynamic properties
of the sample inside the phase boundary \cite{sar}.  Besides, in a
large sample it is not possible to isolate the individual giant
vortex states because the sample encloses a large amount of flux and
a large number of different giant vortex states are equally well
formed at the same applied field.

Recent developments in nano-fabrication techniques now allow us to
reach the limit of isolating these individual giant vortex states in
mesoscopic samples.
We expect the thermodynamic properties of these
isolated giant vortex states (that can be designated an
angular momentum quantum number
$L$) to be completely different from each
other and also from the bulk state. Recently, the phase boundary
\cite{mos,zwe,fom} and magnetization \cite{bui,gei,sch,pal,akk,avi}
of these isolated $L$ states have received some attention. 
Magnetization in superconductors is due to a gauge dependent
equilibrium current and thus it is different in origin from
magnetization of magnetic materials where the ordering of individual
dipole moments, their dynamics and their internal energy scale
determine the magnetization. Heat capacity in superconductors is
determined by the thermal excitations of quasiparticles and therefore,
it can eventually lead to a better understanding of the microscopic
dynamics. For the first time we present a quantitative study of the
heat capacity of the isolated giant vortex states, a quantity which
is accessible in an experiment.  We show that at small fields there
is a discontinuous jump in the heat capacity for $L$=0 state at the
transition temperature, a typical feature of second order phase
transition, but at higher fields this turns into
a continuous change.  The line shape shows drastic changes
during this crossover which can be interpreted as a crossover from
bulk-like to true mesoscopic behavior.  On the other hand the higher
$L$ states show no discontinuity in heat capacity 
at all fields. Quantitative
calculations strongly suggest that these features can be observed
with present day experimental set-ups. Besides, we
show that the paramagnetic
Meissner effect (irreversible)
\cite{gei2} can be indirectly but unambiguously
observed through the heat capacity measurements.

In bulk samples Ginzburg-Landau (GL) theory is valid only close to
the normal-superconductor transition when the gap at the Fermi
surface approaches zero.  So the jump in the specific heat of bulk
samples at $T_c$ agrees with the GL theory with high accuracy for some
metals \cite{boo}.  Well below the transition the observed
specific heat
scales, however, exponentially with temperature, signifying a unique
energy scale associated with excitations and dynamics of
quasiparticles in the system that originates from the presence of a
gap at the Fermi surface, and this does not come out in the GL
theory.
In the giant vortex
state, the GL theory gives the correct magnetization down to 0.33$T_c$
and very small fields \cite{pal,akk}. The giant vortex states are
current carrying edge states
because of which they are sometimes referred to as
``gapless superconducting states'' (for details we refer to
\cite{boo}).
In the absence of a gap the heat
capacity will have a power law behavior below $T_c$
according to the
predictions made here based on the GL theory. The absence
of gap can also be the reason
why the GL theory correctly describes the magnetization at low
temperatures \cite{com2}.

We consider superconducting disks with radius $R$ and thickness $d$
immersed in an insulating medium. Taking the axis of the disk
to be parallel to the z direction and keeping in mind the boundary
condition at $z=\pm d/2$, one can expand the order parameter
as a Fourier series $\psi(z,r)=\Sigma_k cos({k \pi z \over d})
\psi_k(r)$. Unlike in bulk samples $\psi$ is coordinate
dependent in the x-y plane. Substituting in the GL equations
one can check that for $(\pi \xi /d)^2>>1$, the dominant contribution
comes from $\psi_{k=0}$, which naturally means that if the
thickness of the disk is less than the coherence length then
the order parameter cannot vary in the z direction. 
Therefore we may assume a uniform
order parameter in the z direction and as a consequence the
order parameter and currents become two dimensional with
an effective penetration length that increases
with thickness as $\lambda_{eff}=\lambda^2/d$ \cite{sch2}.  
The flux expulsion from the disk becomes negligible when $\lambda_{eff}
\approx R$. In that case the system is quantitatively described by
the GL equation \cite{com}, which is generally accepted \cite{mos,gei}
as a Schr{\"o}dinger like quantum mechanical equation 
\begin{equation}
(-i \vec \nabla -\vec A)^2\Psi=(1-T)\Psi(1-|\Psi|^2),
\end{equation}
that correctly describes the center of mass of the Cooper pairs.
Here $T$ is temperature in units of $T_c(0)$ of bulk
samples. Distance is measured in units of $\xi(0)$,
the coherence length at zero temperature for bulk samples. 
The order parameter is measured in
$\Psi_0(T)$, the finite temperature order parameter of the bulk
sample. The vector potential is
measured in $c\hbar/2e\xi(0)$ and the
magnetic field in  $H_{c2}=c\hbar/e\xi^2(0)$. Our choice of units is
the same as that in Ref. \cite{mae}. On the disk surface we require
that the normal component of the current density is zero, which gives
\begin{equation}
(-i \vec \nabla -\vec A)_n \Psi=0.
\end{equation}
Rescaling the lengths ($R \rightarrow R\sqrt{(1-T)}$) and fields ($H
\rightarrow H/(1-T)$) we rewrite equation (1) in the  following form:
$(-i \vec \nabla -\vec A)^2 \Psi=\Psi (1-|\Psi|^2).$
For the bulk samples the gradient term can be neglected \cite{boo}
and in mesoscopic samples close to the normal superconductor
transition the $|\Psi|^2$ term can be neglected \cite{mos,zwe}. We
show that in the regime of our interest both the terms can be solved
exactly in a semi-analytical way.

The difference of the free energy $G$ between the superconducting and
the normal state, measured  in $H_c^2(0)V/8\pi$, can be expressed
through the integral
\begin{equation}
\label{free}
G=\int \left( 2(1-T)(\vec A-\vec A_0).\vec j-(1-T)^2
|\Psi|^4\right)d\vec r/V,
\end{equation} 
over the disk volume $V=\pi R^2d$, where $\vec A_0=Hr/2 \vec e_{\phi}$
is the external vector potential, and $\vec j=(\Psi^*\vec \nabla
\Psi-\Psi \vec\nabla\Psi^*)/2i -|\Psi |^2\vec A$ is the dimensionless
supercurrent.  The heat capacity difference between the
superconducting and normal states in units of $C_0=H_c(0)^2V /(8 \pi
T_c)$  is defined as $C=-T{d^2G\over dT^2}$. The bulk limit of equation
(3) is $G=H_c(T)^2V/8\pi$ \cite{boo} which yields a specific heat
jump of 2 in units of $C_0$; whereas it can be 
accurately measured to an order of magnitude smaller values at lower
temperatures. As explained before for $d<\xi$ equation (1) becomes two
dimensional \cite{sch2}, i.e.,
\begin{equation}
\left(-i\vec \nabla_{2D} -\vec A\right)^2\Psi=
\Psi (1-|\Psi|^2).
\end{equation}

We write the general solution as a truncated series in the solutions
of the Linearized Ginzburg-Landau (LGL) equation (which is basically
omitting the $|\Psi|^2$ term in (4))  $i.e.,$ $\Psi=A_0^{1/2}
\Gamma_0(\vec r) +  A_L^{1/2} \Gamma_L(\vec r)exp(iL\theta)$. Here
$\Gamma_0(\vec r)$ and $\Gamma_L(\vec r)$ are solutions of the LGL
equation with  eigen energies $\lambda_0$ and $\lambda_L$, $L$ being any
non-zero integer giving the angular momentum quantum number. Hence,
\begin{equation}
\lambda_L=(1+2 \nu) {\phi\over R^2} - 1,
\end{equation}
where $\nu$ is to be determined from the solution of the following
equation.
\begin{equation}
(L-{\phi\over 2})F(-\nu,L+1,{\phi\over 2}) - {\nu \phi \over L+1} 
F(-\nu+1,L+2,{\phi\over 2})=0,
\end{equation}
where $\phi=HR^2$ is the flux through the disk in units of flux
quantum $hc/e$. $F(a,c,y)$ is the Kummer function, and finally
\begin{equation}
\Gamma_L(r)=r^L exp(-{Hr^2 \over 4})F(-\nu,L+1,{Hr^2\over 2}).
\end{equation}
Although we retain only two terms in the series we will show that
this is not an approximation in the regime of our interest. Now
substituting $\Psi$ in equation (4) and simplifying we obtain
\begin{equation}
\lambda_0 A_0 = a_{11} A_0^2 + a_{12} A_0 A_L
\end{equation}
\begin{equation}
\lambda_L A_L = a_{22} A_L^2 + a_{12} A_0 A_L
\end{equation}
where $V a_{11}=\int \Gamma_0^4 dV$, $V a_{22}=\int \Gamma_L^4 dV$
and $V a_{12}=2 \int  \Gamma_0^2 \Gamma_L^2 dV$. We have to solve
equations (8) and (9) to evaluate $A_0$ and $A_L$. Three
possible solutions are
\begin{equation}
(A_0=\lambda_0/a_{11}, \quad A_L=0),
\end{equation}
\begin{equation}
(A_0=0, \quad A_L=\lambda_L/a_{22})
\end{equation}
and
\begin{equation}
(A_0={\lambda_0 a_{22} - \lambda_L a_{12} \over
a_{11}a_{22} - a_{12}^2}
,\quad
A_L={\lambda_L a_{11} - \lambda_0 a_{12} \over
a_{11}a_{22} - a_{12}^2}).
\end{equation}

The solution in (10) corresponds to $L$=0, the solution in (11)
corresponds to $L\ne 0$ giant vortex states and the solution in (12)
corresponds to a general mixed state solution. In (10) and (11),
$a_{ii}$ contains all the correction due to the
non-linear term, $\lambda_i$ being eigenvalue of the LGL equation.
The mixed state
solutions were discussed
earlier \cite{pal2}. As we  increase the
number of terms in the expansion for $\Psi$ the solution (12) becomes
more and more complicated and realistic  but solutions (10) and (11)
remain unchanged and exact. There is a definite parameter regime
where the giant vortex states have a lower free energy than the mixed
vortex states and in this regime equations (10) and (11) give the same
results as the general numerical solutions \cite{sch2}.
The stability analysis of these giant vortex states has been
done analytically from the GL differential equations \cite{sch2} and
numerically from the saddle point analysis of GL free energy
\cite{sch3}, essentially proving their stability. Although
the basic predictions of our work can be tested with just
$L$=0 and the $L$=1 states which are under
all conditions completely symmetric states the higher angular
momentum giant vortex states have been detected experimentally 
\cite{bru}.
Also numerical calculations
\cite{pal} with a larger number of terms in the expansion,
substituted into the GL free energy expression showed that
numerically minimized free energy corresponds to only
one term in the expansion and the rest
are identically zero in certain parameter regimes. This parameter
phase diagram is given in Ref. \cite{pee} and is the regime
of our interest where the solutions (10) and (11) are valid.

In Fig.~1 (a), (b), and (c) we plot $C/C_0$ vs $T/T_c(0)$  at
$\phi/\phi_0$=0.1, 1.0 and 1.5 respectively, for all possible $L$
states for a disk of $R=4.0\xi(0)$. Figs.~1 (b) and (c) are
displaced by 3 and 6, respectively, in the y-direction.
The $L$=0 state is the ground state (shown in Fig.~2)
but in decreasing fields
it is possible to trap the system in the higher $L$ states down to
almost zero field due to the Bean-Livingston barrier
\cite{bea}, when fluxoids are trapped at the center of the
sample and it can carry a large current \cite{gei,deo}. 
In Fig.~1 (a) the
$L$=0 state shows a discontinuity at $T_c$ (we refer to this as
bulk-like behavior), while the higher $L$  states show a continuous
change at their corresponding $T_c$s. 
As the flux is increased in Fig.~1 (b) and (c), 
the discontinuity in the heat capacity
for the $L$=0 state slowly changes to a continuous change,
although there is a striking difference in the line shape of the
$L$=0 state from those of the $L\ne$0 states.
For high enough fields the line
shape of the $L$=0 state will be the same as that of the $L\ne 0$
states as will be shown in Fig.~3.
A discontinuity in
heat capacity is a
characteristic feature of second order phase transition that can
be observed in bulk samples. In finite samples the discontinuity
is replaced by a continuous change.
Symmetry breaking transitions, like solid-liquid melting
transition, that are, however, first order transitions associated
with a divergence in specific heat in the bulk, give a Gaussian
curve in clusters \cite{man}. So the $L$=0 state in Fig.~1 (b)
and (c) exhibits mesoscopic effect but not in Fig.~1 (a).
In the $L$=0
state the Cooper pairs condense into a zero momentum state ($L$ being
its angular momentum) just as bulk superconductivity is macroscopic
condensation to the zero momentum state. Then in the
absence of fields the boundary condition
(2) for the currents at the boundary becomes irrelevant. Thus we
get bulk-like behavior which persists at small but finite fields
even though for infinitesimally small fields the effect of the
boundary conditions set in.
The higher $L$ states 
show no discontinuity in the heat capacity at
all fields. 
This line
shape for the $L$=0 state in Figs.~1 (b) and (c)
is intermediate between mesoscopic
and macroscopic limits and would be an interesting
feature to observe experimentally.

In Fig.~2 we plot $G/G_0$ $(G_0=0.4H_c^2V/(8\pi)$ vs $\phi
/ \phi_0$ for a
disk of $R=4.0\xi(T)$  in dotted lines for different $L$ values. The
solid lines show $(C/C_0)_{peak}$, the magnitude of the peak
value in $C/C_0$ curves at corresponding fields for the different
$L$ states. 
The dashed curve would be $(C/C_0)_{peak}$ for a bulk
sample at zero field.
The solid curves for any $L$ appear  approximately as
a mirror reflection of the dotted curves at the x-axis. The
proportions of the solid and dashed curves are different
in the y-direction because of which the
crossings between the $L$ states in the solid curves occur at the
slightly different fields compared to that in the dotted curves. 
But the minimum in the dotted curve for a particular $L$ is exactly
at the same field of the maximum in the solid curve for the
same $L$. This can be also argued from equation (3).
Normally magnetic field in a bulk superconductor decreases 
the Cooper pair
density and increases the free energy due to loss in
the condensation energy associated with the breaking of
Cooper pairs. But mesoscopic samples
have a regime for each $L$ state
where the free energy
decreases with increase in field (this is nothing but
paramagnetic Meissner effect \cite{gei2} because magnetization
is related to the flux derivative of free energy) as can be seen
from the dotted curves in Fig.~2. In this regime the
Cooper pair density ($|\Psi|^2$) also increase
with increase in magnetic field unlike that
in bulk samples. For example in the inset to Fig.~2,
we consider the $L$=2 state in a regime where it shows the
paramagnetic Meissner effect. The solid curve gives the
Cooper pair density profile along a line starting from the
center of the disk and ending at the boundary
of it at an applied flux of $2\phi_0$. The dotted curve
gives the same at $4\phi_0$. Hence the density at
every point can be enhanced by increasing the field
and as a consequence free energy decreases with increase
in field.
In this regime heat
capacity as well as the peak value of the heat capacity,
at a fixed temperature increases with increase in
magnetic field.
Each $L$ state shows the paramagnetic Meissner
effect at low fields when the field density in the disk
exceeds the applied field. Field density decreases linearly
with the applied field \cite{deo}
and at higher fields in the diamagnetic
regime the field density inside the disk is less than the applied
field. In this monotonous field dependence the magnetic energy
is proportional to the square of the field expelled (being always
positive) and hence the free energy
minimizes when Cooper pair density maximizes and field
density is the same outside as well as inside the sample.
In fact any observable quantity (like heat capacity)
that depends on the density
can identify the minima of the free energy
and hence the regime where it
decreases with increase in field or the regime of paramagnetic
Meissner effect.
Hence at any fixed temperature, if
the heat capacity of a mesoscopic sample shows enhancement
with increasing field, it can be taken as an
indirect but unambiguous observation of the paramagnetic
Meissner effect (irreversible).
Such an unambiguous observation is not possible by a direct
magnetization measurement as has been clearly shown in Fig.~11
in Ref. \cite{deo}:
while the sample is always diamagnetic the observed
magnetization is most of the time paramagnetic.
The solid curve for the
$L$=0 state at small
flux shows a sudden jump that is related to the
crossover from bulk-like to true
mesoscopic behavior.

For the higher $L$ states the difference in $(C/C_0)_{peak}$ 
in Fig.~2 becomes
small for consecutive $L$ states but this will not create accuracy
problems in measuring $(C/C_0)$ or $(C/C_0)_{peak}$ of isolated $L$
states up to considerably large $L$ values, because the transition
temperatures of the different $L$ states are different \cite{mos}.
Therefore, the $(C/C_0)$ for consecutive $L$ states peak at different
temperatures and there can be a large difference between $(C/C_0)$ as
well as $(C/C_0)_{peak}$  of consecutive $L$ states at particular
temperatures which will help to isolate them experimentally. For
example, we show the $(C/C_0)$ for all possible $L$ states at 
$\phi/\phi_0$=11.4 in Fig.~3. 
The ground state at this field is the $L$=3 state (shown in 
Fig.~2). The difference in $(C/C_0)$ for consecutive $L$ states at
certain temperatures is accentuated by arrows and it can be much
larger than the corresponding differences in $(C/C_0)_{peak}$. The
figure also shows that at large fields there is no noticeable line
shape difference between the $L$=0 and $L\ne 0$ giant vortex states.

Metastability can create some problems
in an experimental situation but one faces the same problem
in magnetization experiments. However, by systematically
sweeping up and
down the field one can perform measurements in almost
the entire relevant part of phase space. One can just measure
heat capacity as a function of magnetic field and in turn
obtain information about heat capacity as a function
of temperature. Ref. \cite{gei2} also proves that by varying magnetic
field at constant temperature or by varying temperature
at a constant field one can arrive at the same stable state.
So keeping the experimental set up unchanged one has to
replace the detector by a calorimeter.

In summary, we have shown that superconducting disks can give us an
opportunity to observe a crossover of a
second order phase transition in
bulk system to such a transition in a finite system, where the
discontinuity in heat capacity at the transition temperature is
gradually smoothed out, accompanied by drastic changes
in the line shape. The
giant vortex states with non-zero angular momentum
can exhibit finite size effects
like a continuous change in the heat capacity at the transition
temperature. Heat capacity measurements 
can also lead to an unambiguous verification of the
existence of paramagnetic Meissner effect  (irreversible)
in mesoscopic samples.
Besides, we show that the GL theory can be sufficiently
simplified to incorporate mesoscopic corrections to
heat capacity because it retains accurately the spatial
variations of order parameter that is determined by
the coherence length and quiet independent of the
penetration length for $\lambda_{eff}>R$.

This work is supported by the Academy of Finland and was done
during PSD's stay at the University of Jyv{\"a}skyl{\"a}.
One of us (PSD) acknowledges several useful discussions with Prof. F.
M. Peeters and Prof. V. A. Schweigert.

\centerline{Figure captions}

\noindent Fig.~1 (a) $C/C_0$ versus $T/T_c(0)$ for a disk of radius
$R=4 \xi(0)$ at a flux $\phi/\phi_0$=0.1 for all possible $L$ states.
(b) The same as (a) but $\phi/\phi_0$=1.0. (c) $\phi/\phi_0$=1.5.

\noindent Fig.~2. The solid curves show $(C/C_0)_{peak}$ versus 
$\phi/\phi_0$ for all possible $L$ states. The dotted curves
show $(G/G_0)$ versus 
$\phi/\phi_0$ for all possible $L$ states. The first few $L$
states are labeled with their corresponding $L$ values.
Radius $R$ of the disk=4.0$\xi(T)$. The dashed curve gives
$(C/C_0)_{peak}$ for a bulk sample at zero field having the same
coherence length. The inset shows the density profile along
a radial line of the disk at two different fields.
$C_0$ and $G_0$, defined in text, contains all sample parameters.
Thus the figure holds for all samples and at any temperature,
the radius being four times the coherence length at that
temperature. The radius in microns has to be thus
properly chosen using the equations defined immediately below equation 2.

\noindent Fig.~3. $C/C_0$ versus $T/T_c(0)$ for a disk of radius $R=4
\xi(0)$ at a flux $\phi/\phi_0$=11.4 for all possible $L$ states.

\begin{thebibliography}{99}
\bibitem{com2} The GL equations can be derived from the BCS Hamiltonian
under the single approximation that the gap is vanishingly small.
\bibitem{com} Otherwise as shown 
in \cite{pal} $\vec A$ in equation (1) can be
replaced by the one corresponding to $<\vec B>$, the average
induced magnetic field,
which does not change the
qualitative or quantitative features, except that the transitions
between different $L$ states occur at slightly higher fields.
\end{thebibliography}
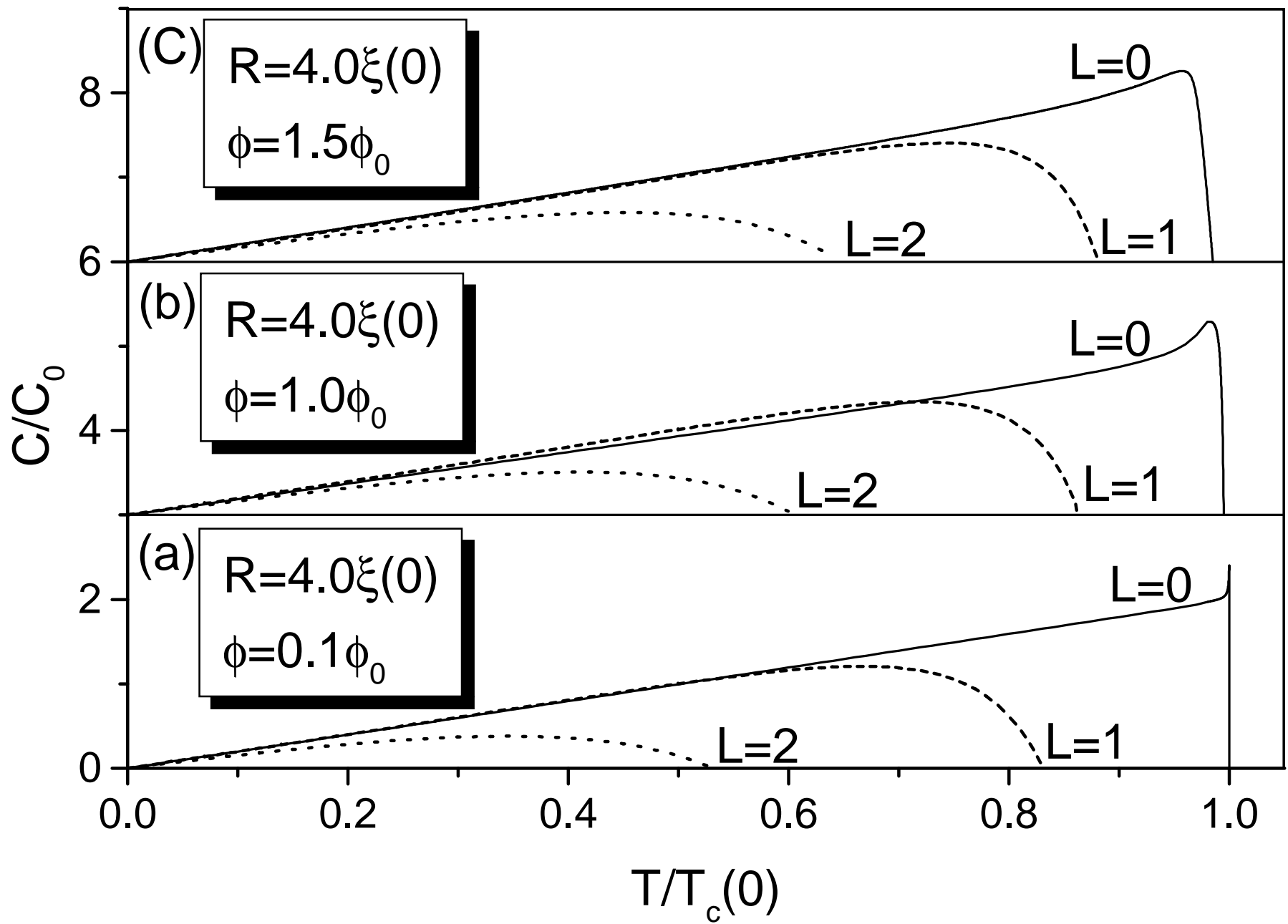
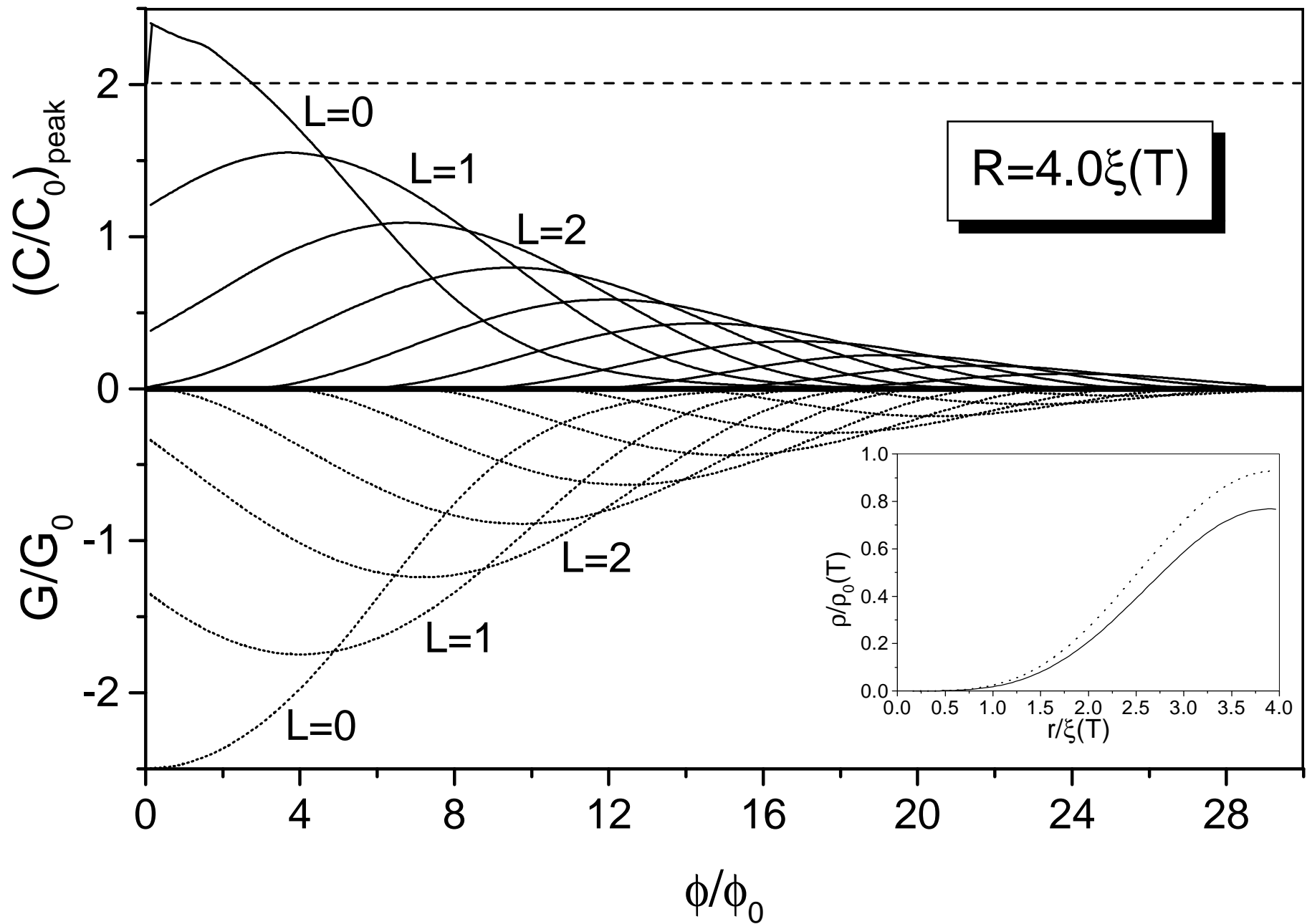
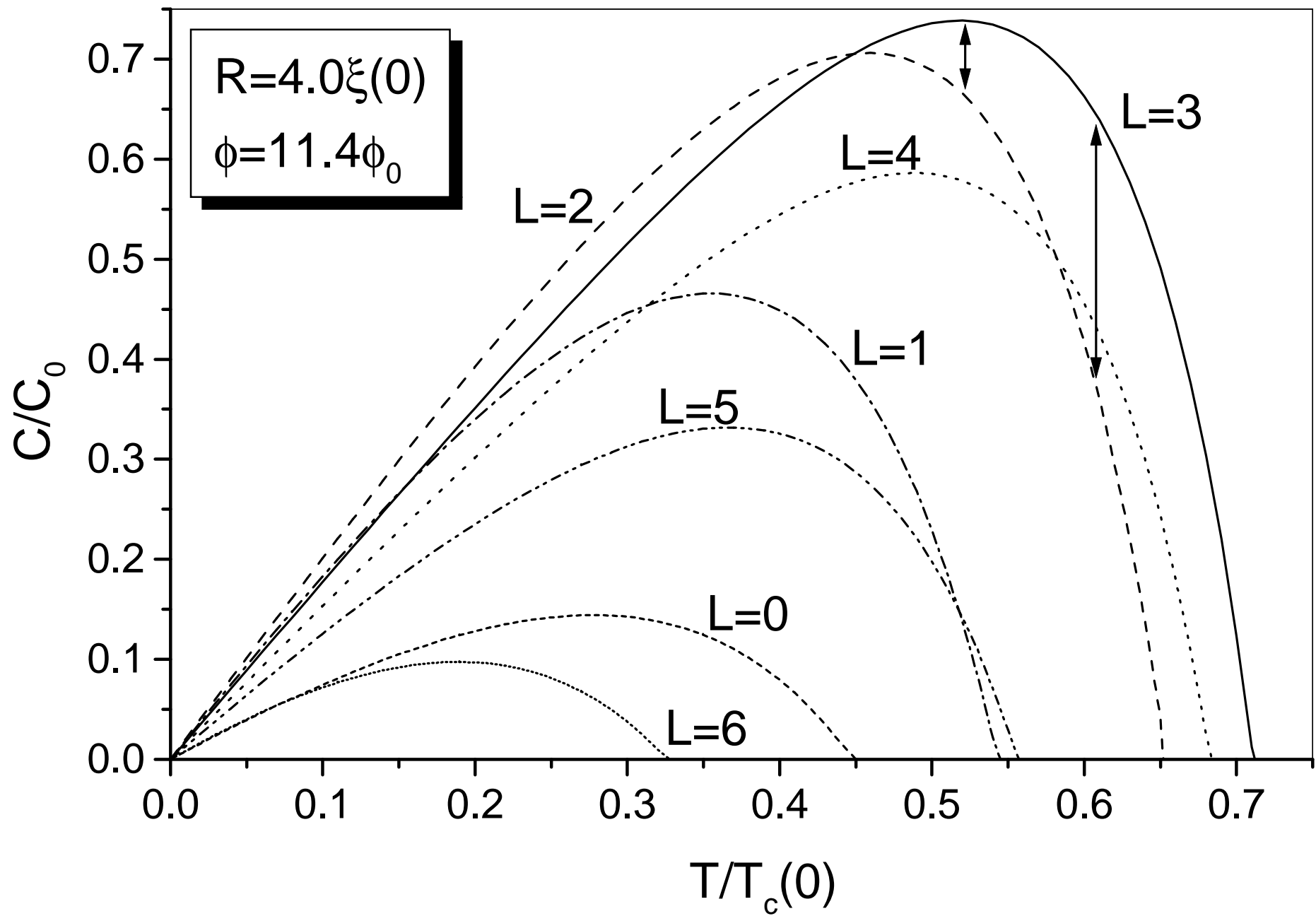
\end{document}